\documentclass[3p,times,procedia]{elsarticle}
\usepackage{nupha_ecrc} 
\usepackage{float}
\usepackage{multicol}
\usepackage{lineno}

\volume{00}

\firstpage{1}

\journalname{Nuclear Physics A}

\runauth{Kishora Nayak}

\jid{nupha}

\jnltitlelogo{Nuclear Physics A}

\usepackage{amssymb}

\usepackage[figuresright]{rotating}

\newcommand{\pt}{\it{p}_{\rm T}}

\begin{document}

\begin{frontmatter}



\author{Kishora Nayak (for the STAR Collaboration)}
\address{Key Laboratory of Quark and Lepton Physics (MOE) and Institute of Particle Physics, \\ Central China Normal University, Wuhan 430079, China}
 \ead{k.nayak1234@gmail.com}

\dochead{XXVIIIth International Conference on Ultrarelativistic Nucleus-Nucleus Collisions\\ (Quark Matter 2019)}

\title{Directed and elliptic flow of identified hadrons, high-$\it{p}_{\rm T}$ charged hadrons and light nuclei
in Au+Au collisions at STAR}


\author{}

\address{}

\begin{abstract}
In this proceeding, we present the new precise directed flow $v_{1}$, measurement of $\pi$, $K$, $p$ and $\phi$ in Au+Au 
collisions at $\sqrt{s_{\rm NN}}$ = 27, 54.4 GeV and deuteron at $\sqrt{s_{\rm NN}}$ = 7.7, 11.5, 14.5 and 19.6 GeV. The 
first measurement of pseudorapidity and centrality dependence of the $v_{1}$ of high-$p_{T}$ ($>$5 GeV/$c$) charged 
hadrons in Au+Au collisions at $\sqrt{s_{\rm NN}}$ =  200 GeV are reported. The elliptic flow $v_{2}$, of identified 
hadrons ($\pi$, $K$, $p$, $\phi$, $K^{0}_{S}$, $\Lambda$, $\Xi$, $\Omega$) in Au+Au collisions at  $\sqrt{s_{\rm NN}}$ 
= 54.4 GeV are also presented. The results are compared with model calculations to discuss the physics implications.

\end{abstract}


\end{frontmatter}


\section{Introduction}
One of the important goal of the STAR beam energy scan program at RHIC is to understand the QCD phase 
diagram of the strongly interacting matter produced in the relativistic heavy-ion collisions. Collective flow 
phenomena are sensitive probes to characterize the properties of the produced QCD matter~\cite{qcd}. The 
measured flow observables are compared with model calculations to constrain the Equation of State (EoS) and 
to understand the QCD phenomenon.  The first and second order harmonic of the Fourier expansion of the 
emitted particles in momentum space are characterized as directed flow ($v_{1}$) and elliptic flow ($v_{2}$), 
respectively~\cite{flowDef}. The rapidity-odd component of $v_{1}$ and $v_{2}$ are sensitive probes of the bulk 
to study the collective dynamics in the early stage of the collisions. The transport and hydrodynamic model 
calculation suggested that the negative $v_{1}$-slope for baryons as a function of beam energy is a signature 
of first order phase transition~\cite{hydro1, amptV1, jam}. The $v_{1}$ of high-$\pt$ charged hadrons 
measurement are expected to give valuable constraints on the initial longitudinal distribution of the fireball and
also give idea about the path length-dependent energy loss of partons.

The thermalized QCD matter formed in heavy-ion collisions is tilted in the reaction plane as a function of rapidity, 
while the production profile of partons from hard scatterings is symmetric in rapidity~\cite{third,third1}. This leads 
to a rapidity-odd directed flow ($v_{1}$) for high-$p_{T}$ hadrons and can provide valuable constraints on the 
initial longitudinal distribution of the fireball as well as the path length-dependent energy loss of partons. 
Hydrodynamic models suggest that double sign change in the beam energy dependence of $v_1$ slope as 
a function of rapidity $y$, d$v_1$/d$y$, around $y$ = 0 for net baryon is a signature of the first-order phase 
transition~\cite{firstOrder}. The light nuclei and strange hadrons might be more sensitive to the early EoS 
because of their heavy masses and smaller hadronic interaction cross section, respectively. Due to the different 
sensitivity of strange particles to hadronic phases, the mass ordering of elliptic flow ($v_{2}$) is expected to be 
violated between proton and $\phi$ meson in the low-$p_{\rm T}$ range ($p_{\rm T}<$1.0 GeV/$c$)~\cite{rescat1,rescat2}.

\section{Elliptic flow of identified hadrons}
\begin{figure*}[!h]
\begin{center}
\includegraphics[scale=0.31]{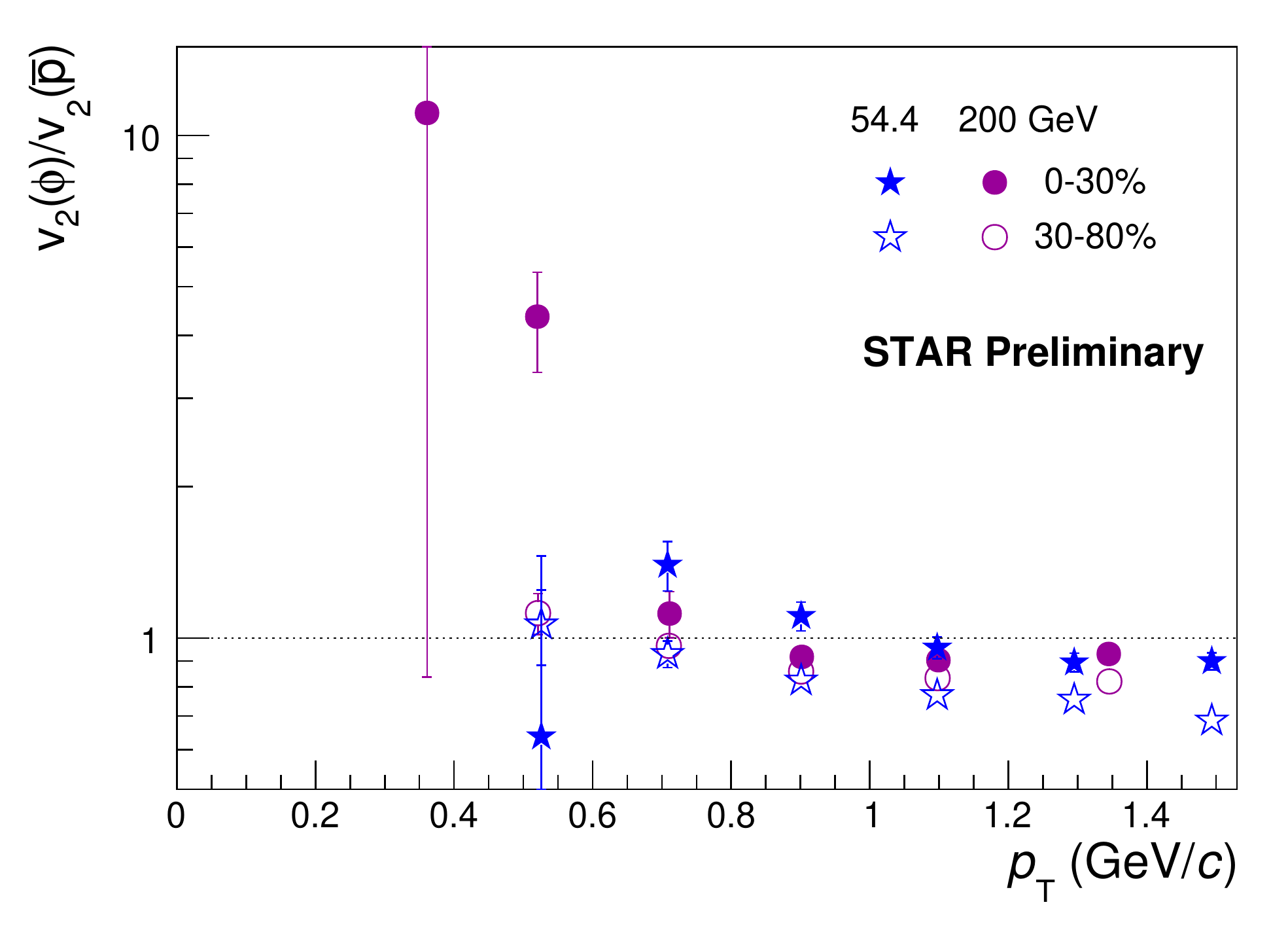}
\includegraphics[scale=0.28]{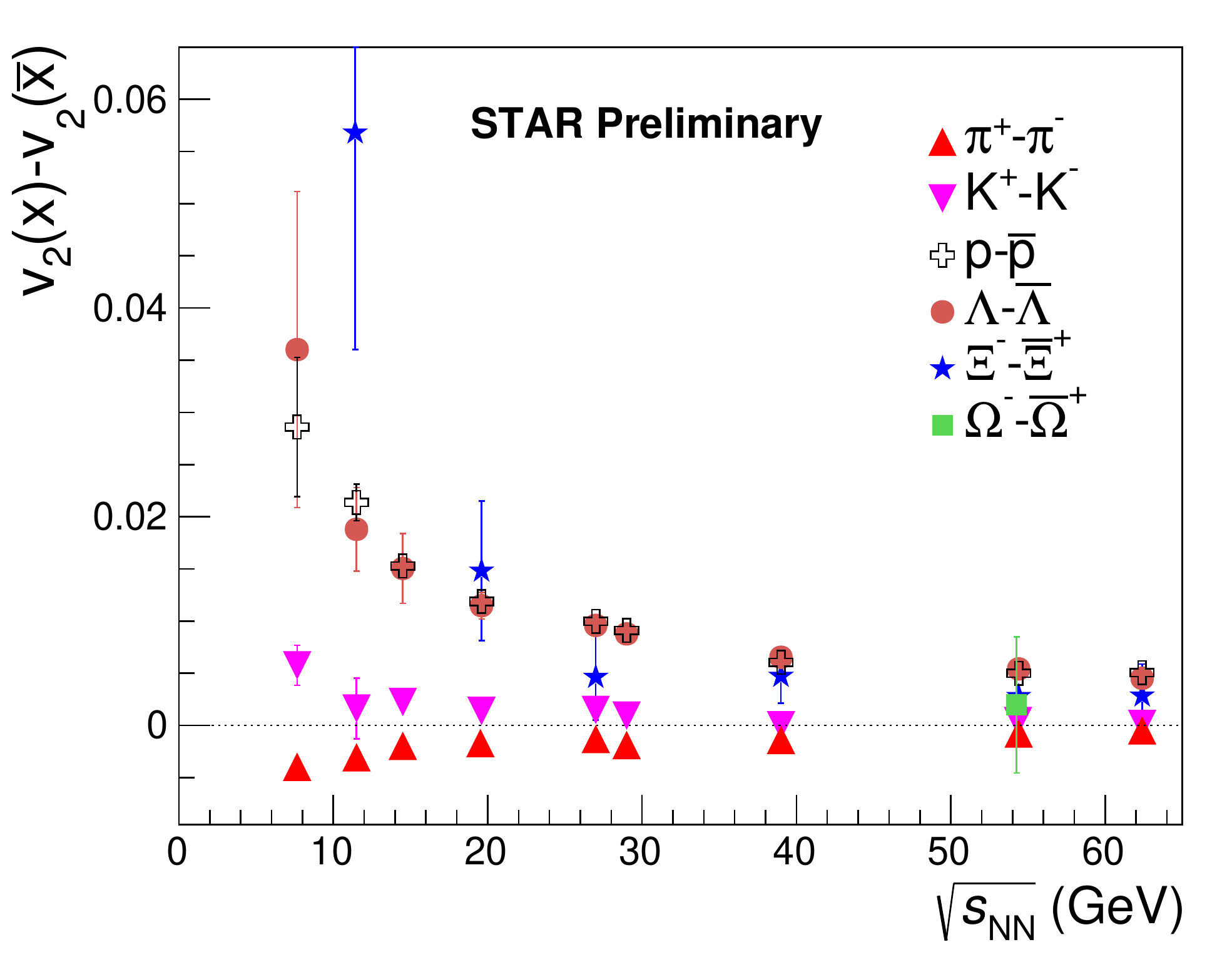}
\caption{(Left) $\it{p}_{\rm T}$ dependent ratio of $v_{2} (\phi)/v_{2} (\bar{p})$ for 0-30$\%$ and 30-80$\%$ 
centralities in Au+Au collisions at $\sqrt{s_{\rm NN}}$ = 54.4 and 200 GeV. (Right) $v_{2}$ difference between 
the particle and anti-particle as a function of beam energy for minimum bias Au+Au collisions.}
 \label{fig1}
\end{center}
\end{figure*}

The $\pt$ dependent $v_{2}(\phi)/v_{2}(\bar{p})$ ratio for two different centralities in Au+Au collisions at 
$\sqrt{s_{\rm NN}}$ = 54.4 GeV and 200 GeV are shown in Fig.~\ref{fig1} (left)~\cite{starv2}. The mass of $\phi$ 
meson is higher than proton, so the $v_2$ of $\phi$ meson is expected to be smaller than proton (hydrodynamic 
behavior) in the same $\pt$ bin unlike the observation reported here. This violation of mass ordering at low-$\pt$ 
in central collisions is attributed to rescattering effect by the proton in the hadronic medium unlike the $\phi$ 
meson which remains unaffected by the medium~\cite{rescatter, rescatter22}. The model study indicates that with 
increasing hadronic rescatterings the $v_{2}(\phi)/v_{2}(\bar{p})$ ratio increases. This violation of mass ordering is 
also observed to depend on energy and centrality. The $v_{2}$ difference between the particle and anti-particle as 
a function of beam energy is shown in Fig.~\ref{fig1} (right). The difference between baryon and anti-baryon $v_{2}$ 
increases more compare to mesons with decrease in energy. The number of constituent quark (NCQ) scaling violation 
at low energy is observed. The new $v_{2}$ measurement for strange ($K$ and $\Lambda$) and multistarange ($\Omega$ 
and $\Xi$) hadrons at $\sqrt{s_{\rm NN}}$ = 27 and 54.4 GeV also follow the overall energy dependence.  

\section{Directed flow of identified hadrons and light nuclei}
\begin{figure*}[!h]
  \begin{center}
    \includegraphics[scale=0.35]{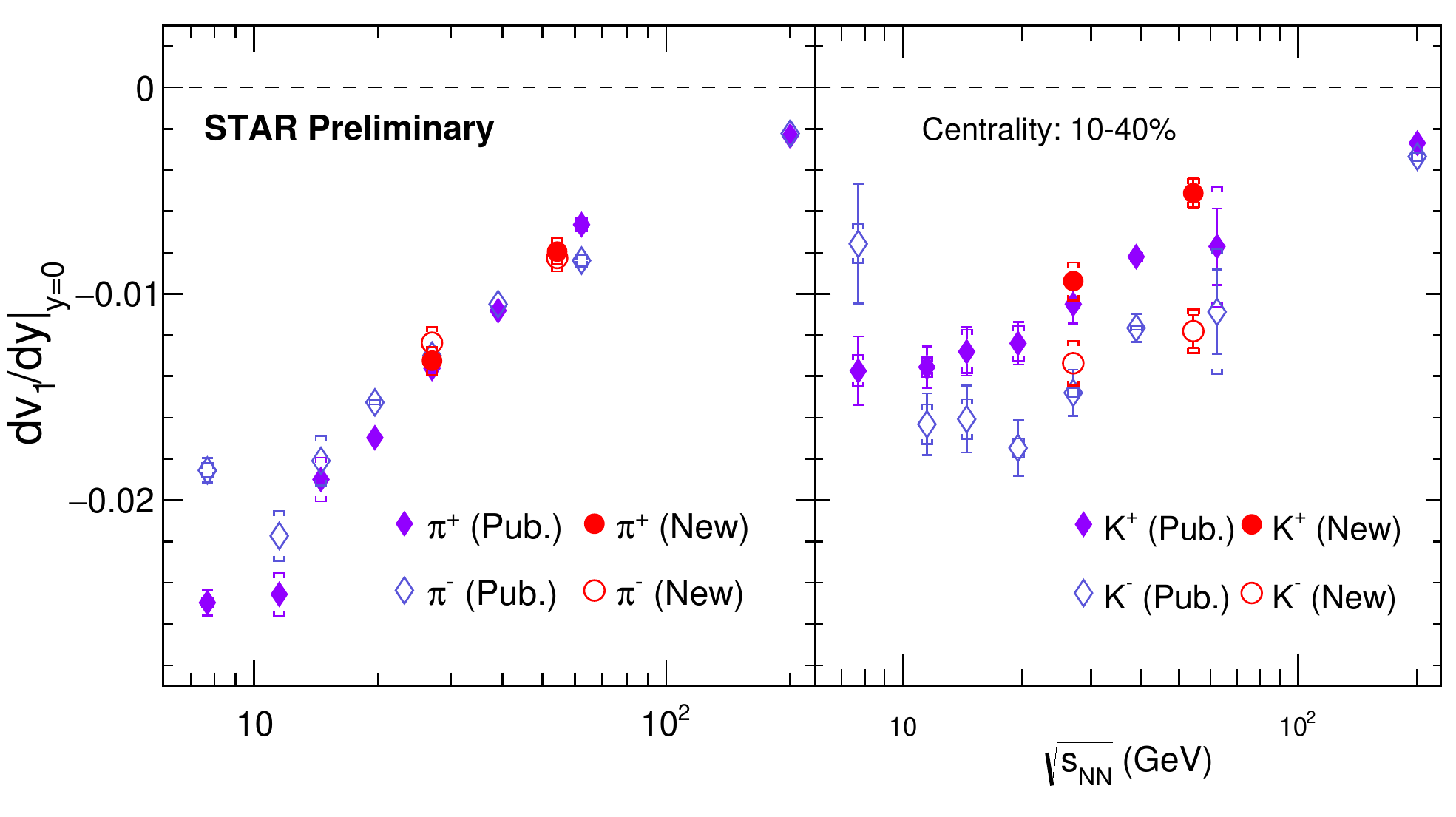}
    \includegraphics[scale=0.28]{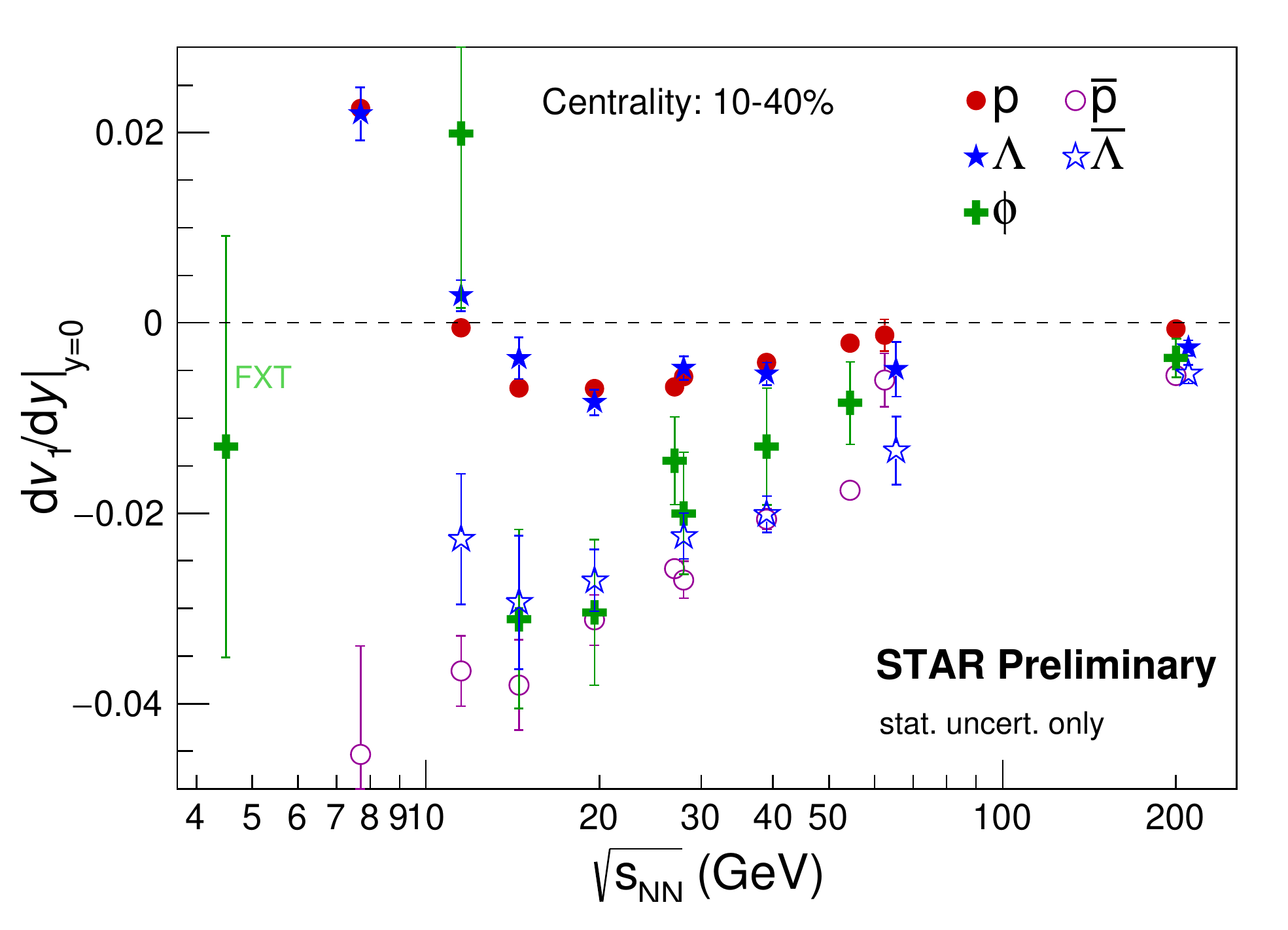}
    \caption{$v_{1}$ slope (d$v_{1}$/dy) at mid rapidity as a function of beam energy in 10-40$\%$ centrality 
      Au+Au collisions for $\pi^{\pm}$ (Left), $K^{\pm}$ (Middle) and $p, \bar{p}, \Lambda,\bar{\Lambda}, \phi$ 
      (Right)~\cite{starv1}. }
    \label{fig2}
  \end{center}
\end{figure*}

The beam energy dependence of $v_1$ slope at mid-rapidity in 10-40$\%$ centrality Au+Au collisions for $\pi^{\pm}$ 
(left), $K^{\pm}$ (middle) and $p, \bar{p}, \Lambda, \bar{\Lambda}, \phi$ (right) are shown in Fig.~\ref{fig2}. The $v_1$ 
slope parameter is extracted by fitting the $v_{1}(y)$ with a linear function over $|y|<$0.8. The slope of anti-baryons 
and mesons are found to be negative for all the measured energies. However, there is a slope change of baryons (proton 
and $\Lambda$) from positive to negative with increase in energy around $\sqrt{s_{\rm NN}}$ = 14.5 GeV, which might 
be an indication of softening of EoS around the same energy as predicted by various hydrodynamics and a transport 
models~\cite{hydro1,jam}. As per the hydrodynamic calculation, a minimum in directed flow has been proposed 
as a signature of a first order phase transition between hadronic matter and Quark-Gluon Plasma phases. The first 
measurement of $\phi$ meson slope from fixed target experiment in Au+Au collisions at $\sqrt{\it{s}_{\rm NN}}$ 
= 4.5 GeV is also reported. 

\begin{figure}[H]
    \begin{center}
      \includegraphics[scale=0.36]{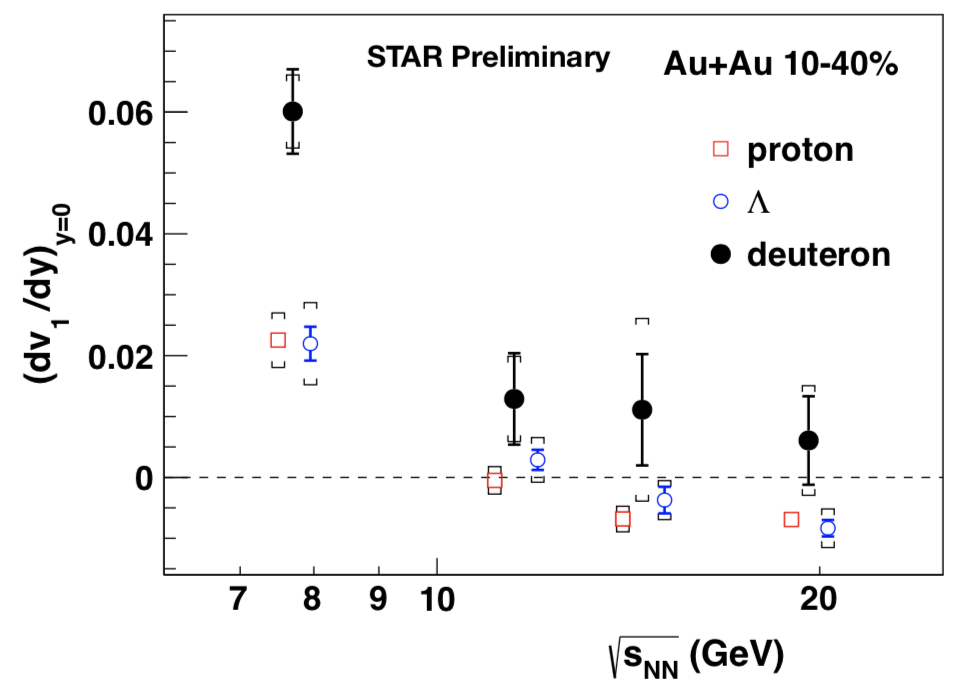}
      \caption{$v_{1}$ slope (d$v_{1}$/dy) at mid rapidity as a function of $\sqrt{s_{\rm NN}}$ in 10-40$\%$ 
        centrality Au+Au collisions for deuteron, $p$ and $\Lambda$.}
      \label{fig3}
    \end{center}
  \end{figure}

\begin{multicols}{2}{
    The light nuclei (deuteron) $v_{1}$ measurement at various collision beam energies  is sensitive to  
    the nuclei formation mechanism. The comparison of deuteron $v_1$ slope with proton and $\Lambda$ for 
    midcentral  (10-40$\%$) Au+Au collisions at $\sqrt{s_{\rm NN}}$ = 7.7-19.6 GeV is shown in Fig.~\ref{fig3}. 
    The slope parameter are extracted by fitting the $v_{1}$ vs rapidity distribution with a linear fit function 
    similar to other light hadrons but the fitting range is $|y|<0.6$. The slope of deuteron is positive or consistent 
    with zero in the measured energy range. At $\sqrt{s_{\rm NN}}$ = 7.7 GeV, the slope of deuteron is positive 
    and more than twice to that of proton or $\Lambda$ unlike the other measured energies.   
  }
  
\end{multicols}

\section{Directed flow of high-$\pt$ charged hadrons}
 In hydrodynamics picture, the QGP fireball represented by soft particles is tilted in rapidity. However, the hard 
 scattering profile is symmetric in nature. This indicates that there is hard-soft asymmetry in the initial state 
which induces negative $v_{1}$ for hard partons~\cite{third,third1}. Similar effect along with the drag by the bulk 
produces a large $v_{1}$ for $D^{0}$ meson~\cite{third1}. The hydrodynamic models successfully explain the 
large $v_1$ slope of $D^{0}$ mesons by suitably choosing the tilt and drag parameters~\cite{third1}. The path-length 
dependence energy loss and constrain to initial condition in longitudinal direction can be studied by measuring 
$v_{1}$ of charged hadrons at high-$\pt$.
 
\begin{figure*}[!h]
\begin{center}
\includegraphics[height=3.6cm, width=5.5cm]{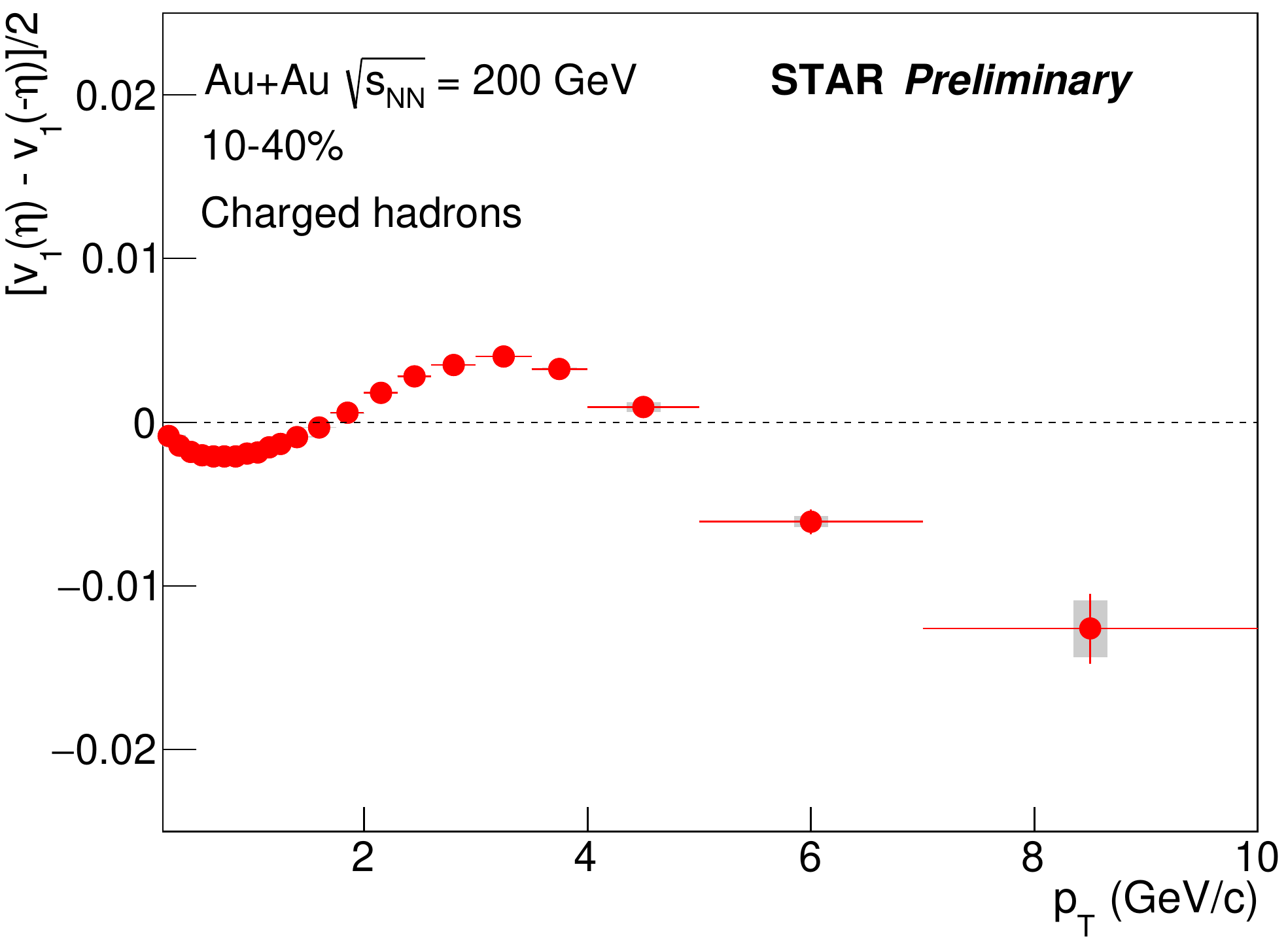}
\includegraphics[height=3.6cm, width=5.5cm]{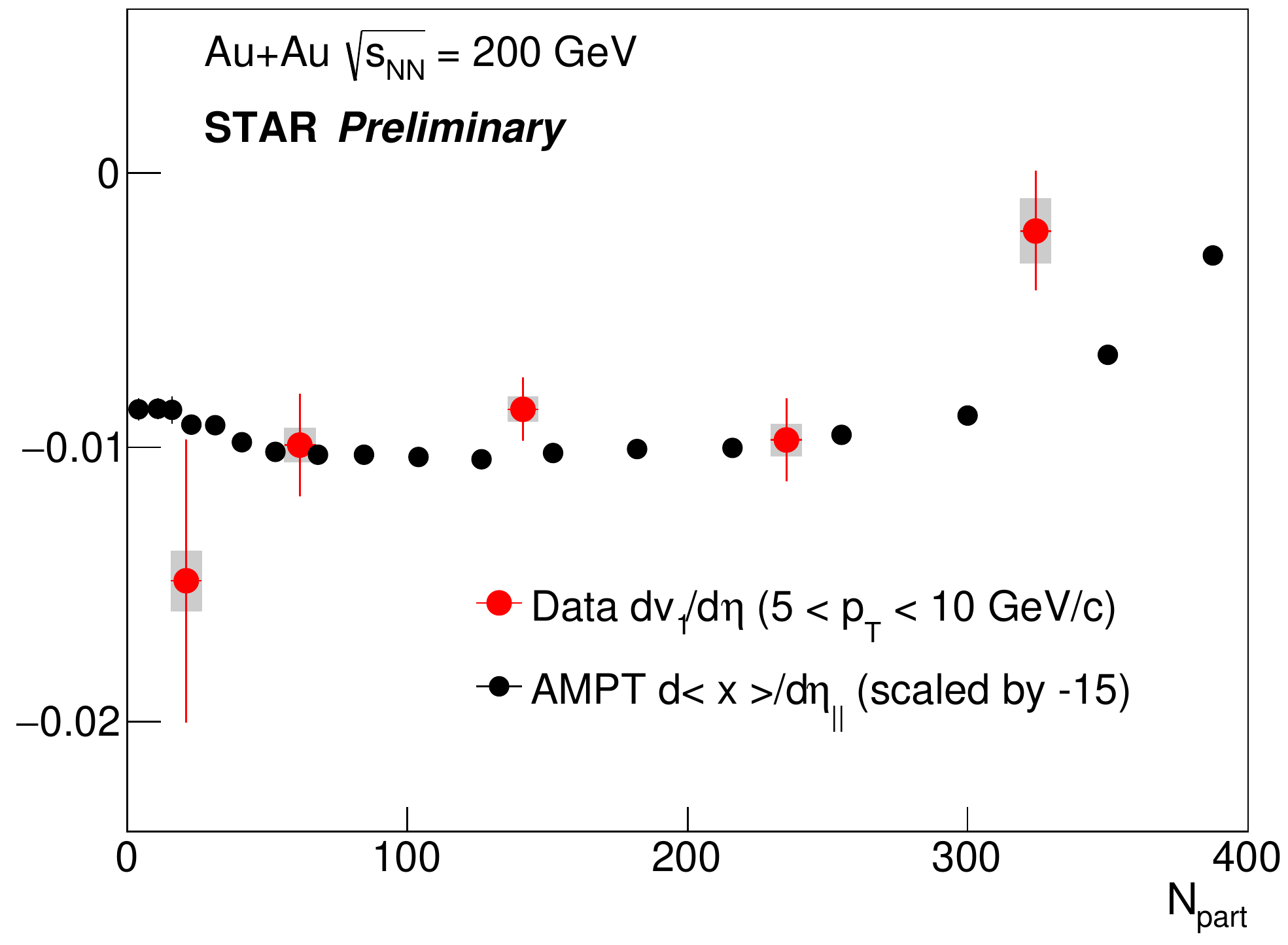}
\caption{(Left) $\pt$ dependent $v_{1}$ of charged hadrons in positive (0.5 $< \eta <$ 1.0) and negative (-1.0 $< \eta <$ -0.5) 
pseudorapidity for 10-40$\%$ centrality, Au+Au collisions at $\sqrt{s_{\rm NN}}$ = 200 GeV. (Right) High-$\pt$ (5-10 GeV/$c$) 
charged hadrons $v_{1}$ slope (d$v_{1}$/d$\eta$) as a function centrality in Au+Au collisions at 
$\sqrt{s_{\rm NN}}$ = 200 GeV. }
 \label{fig4}
\end{center}
\end{figure*}

The $\pt$ dependence average $v_1$ of charged hadrons in the pseudorapidity 0.5 $<$ $|\eta|$ $<$ 1.0  for 10-40$\%$ 
centrality in Au+Au collisions at $\sqrt{s_{\rm NN}}$ = 200 GeV is shown in the Fig.~\ref{fig4} (left). The distribution 
changes sign twice round $\pt$ $\sim$ 1.8 and 5 GeV/$c$. At high-$\pt$ ($\pt$ $>$ 5 GeV/$c$), the $v_{1}$ magnitude 
has large negative value similar to $D^{0}$ meson. This observation suggests the path-length dependent energy loss 
for high-$\pt$ hadrons. The $v_1$ slope of high-$\pt$ (5 $<$ $\pt$ $<$ 10 GeV$c$) charged hadrons as a function of 
$N_{\rm part}$ (or centralities) in Au+Au collisions at $\sqrt{s_{\rm NN}}$ = 200 GeV is shown in the Fig.~\ref{fig4} (right). 
It is also compared with the corresponding AMPT model calculation. The results are scaled by a factor -15 and the negative 
sign to take care the configuration space (AMPT) and momentum space (data) calculation scenario. The AMPT results are 
qualitatively in a good agreement with the data. The slope has weak centrality dependence i.e for peripheral collisions 
the slope is higher compared to midcentral and central collisions. This characteristic is understood as the initial asymmetry 
is expected to be small for the most central collisions as compared to peripheral collisions.   

\section{Conclusion}
\label{conc}     
The recent measurement of identified hadrons $v_{2}$ in Au+Au collisions at $\sqrt{s_{\rm NN}}$ = 27 GeV and 54.4 GeV is 
reported. Mass ordering violation between $v_2(\bar{p})$ and $v_2(\phi)$ in central collisions for $\sqrt{s_{\rm NN}}$ = 54.4 GeV 
and 200 GeV is presented. This violation might be due to the hadronic rescattering of proton in the hadronic phase as predicted 
by various model calculations. The new measurements of $v_{2}$ difference between particles and antiparticles which includes 
strange and multistrange hadrons as a function of beam energy are in good agreement with the earlier measured results. 
The $v_1$ slope of mesons and antibaryons are negative over the measured energy range and mostly the magnitude increases 
with decrease in energy. This negative slope can be understood if we consider the tilt of QGP bulk in rapidity as per the 
hydrodynamic explanation.  The new measurements of $\pi$, K, $p$ and $\phi$ mesons slope in Au+Au collisions at 
$\sqrt{s_{\rm NN}}$ = 27 GeV and 54.4 GeV are in a good agreement with the earlier measurement. The deuteron $v_1$ 
slope for 10-40$\%$ centrality in Au+Au collisions at $\sqrt{s_{\rm NN}}$ = 7.7 - 19.6 GeV is positive or consistent with zero. 
The negative high-$\pt$ charged hadron $v_1$ is consistent with the path-length dependence energy loss picture. The weak 
centrality dependence of $v_1$ slope is a result of hard-soft asymmetry from the initial stage which is supported by the AMPT 
model calculations.  

\section{Acknowledgments}
\label{ack}  
This work is supported in part by the National Natural Science Foundation of China under Grant No. 11890711 and China 
Postdoctoral Science Foundation under Grant No. 2019M662681.





\bibliographystyle{elsarticle-num}
\bibliography{<your-bib-database>}



\end{document}